\documentclass[12pt,notoc]{JHEP3}

\usepackage{amsmath,amssymb,euscript,array,cite}

\input{epsf}
\usepackage{epsfig}

\def\N{{\cal N}}

\def\Dbarslash{\,\,{\raise.15ex\hbox{/}\mkern-12mu {\bar D}}}
\def\Dslash{\,\,{\raise.15ex\hbox{/}\mkern-12mu D}}
\def\delslash{\,\,{\raise.15ex\hbox{/}\mkern-9mu \partial}}
\def\delbarslash{\,\,{\raise.15ex\hbox{/}\mkern-9mu {\bar\partial}}}

\newcommand{\MAT}[1]{\begin{pmatrix} #1\end{pmatrix}}
\newcommand{\EQ}[1]{\begin{equation} #1 \end{equation}}

\newcommand{\SP}[1]{\begin{equation}\begin{split} #1
\end{split}\end{equation}}

\title{Interactions of 
Domain Walls of SUSY Yang-Mills as D-Branes}
\author{Adi Armoni and Timothy J. Hollowood \\
Department of Physics,\\ University of Wales Swansea,\\
Swansea, SA2 8PP, UK.\\
E-mail: {\tt a.armoni,t.hollowood@swan.ac.uk}}
\preprint{SWAT/06/456}
\abstract{Domain walls in supersymmetric Yang-Mills are BPS
configurations which preserve two supercharges of the parent
theory and so their tensions are known exactly. On the other hand, 
they have been described as D-branes for the confining string. 
This leads to a description of their collective dynamics in terms of a
$2+1$-dimensional gauge theory with two supersymmetries 
and a Chern-Simons term. We show that this open string description can
capture the qualitative behaviour of the forces between the domain
walls for an arbitrary configuration of $n$ walls at leading order in
$1/N$, extending earlier calculations for two walls. The potential admits a 
supersymmetric bound state when the $n$ walls are all coincident 
and asymptotes to a constant at large separation 
with an $n$ dependence which agrees perfectly with the exact
tension formula.
}

\begin{document}

\section{Introduction}

Supersymmetric gauge theories are fascinating because they have all
the physical properties of QCD itself in a situation where some
quantities, the holomorphic ones, can be calculated exactly. For
instance, this includes the condensates of lowest components of
chiral operators like the gluino condensate. This leads to an exact
description (modulo some caveats) of the vacuum structure of the
theory. For gauge group $SU(N)$ there are $N$ discrete vacua for which
\EQ{
\langle\lambda\lambda\rangle_ j=N\Lambda^3 \exp \left (i\frac{2\pi j}{N} \right ) \ ,
}
$j=0,1,\ldots,N-1$. However if one is completely honest, impressive
though this
holomorphic data is it only scratches the surface of the gauge
theory. One should strive for more. In theories with extended
supersymmetry, one can make exact statements about BPS states which
preserve a certain proportion of the supersymmetry, usually one
half. The mass of such states is then determined by a central charge
in the supersymmetry algebra. In theories with minimal supersymmetry
in four dimensions, however, BPS particles states do not exist, since there is
no appropriate central charge in the supersymmetry algebra. Nevertheless,
there is a tensorial central charge which allows co-dimension objects 
to be BPS. Dvali and Shifman \cite{Dvali:1996xe} showed
that the domain walls which separate the discrete vacua are such BPS
objects and the supersymmetry algebra yields an exact formula for the
tension. For the wall which separates the $j^\text{th}$ and
$j+n^\text{th}$ vacua (the label is to be understood modulo $N$) the tension is
\EQ{
T_n=\frac{N^2\Lambda^3}{4\pi^2}\sin\frac{\pi n}{N}\ .
\label{tk}
}
$n=1,\ldots,N-1$. It is remarkable that this formula includes all the
quantum effects and is exact. It follows from this simple formula that
there must be forces between domain walls. Imagine a configuration of
two parallel plane domain walls with vacua $j$ to the left,
$j+n_1+n_2$ to the right, and $j+n_1$ in between. Since
$T_{n_1}+T_{n_2}>T_{n_1+n_2}$ it is clearly energetically
advantageous for the domain walls to move together and squeeze
away the vacuum in the middle to form a bound state. Of course the
above argument doesn't tell us what the force is, just that the
potential must rise from zero, since the bound state is BPS, and must
asymptote to a constant to account for the binding energy per unit
area $\Delta T_n=nT_1-T_n$. It must be possible to interpret the force
between the walls in terms of 
the exchange of the particle states in the theory, in this
case from the tower of glueballs. This description should be good at
large distances, compared with $\Lambda^{-1}$, and the potential will have
the behaviour \cite{Armoni:2003jk}
\EQ{
V(X)\underset{X\to\infty}=V_0+\sum_iC_ie^{-M_i X}\ .
}
For short distances, $X\ll|\Lambda|$ and the sum over the tower of glueballs
will be less useful. {\it A priori\/}, since the domain walls are
non-perturbative objects, it is not clear how to describe the short
distance potential.

If we were working strictly within the confines of 
field theory, then this would
probably be the end of the story. However, the gauge theory can be
engineered in string theory and these constructions imply 
that the domain walls are precisely D-branes for the confining
string \cite{Witten:1997ep} (see also \cite{Tong:2005nf} for a recent
discussion). 
This means that the forces between the walls can--at least in
principle---be determined by considering open string interactions between
the walls. In particular, this description should be valid at small
separations. The light degrees-of-freedom of $n$ walls (the approximate
moduli) are described by a $U(n)$ gauge theory on the walls with a
single (real) adjoint scalar field describing the fluctuations of the
walls in the transverse direction. The action describing these light
fields will be some very complication Born-Infeld type theory
interacting with higher mass string states. However, Acharya and Vafa
suggested that the truncation of this theory to the terms most
relevant at low energy was 
a supersymmetric Yang-Mills-Chern-Simons theory \cite{Acharya:2001dz} :
\SP{
{\cal
  L}&=\frac1{2g^2}\text{Tr}\Big(-(D_i\phi)^2-\tfrac12(F_{ij})^2-i
\chi\Dslash\chi-i\psi\Dslash\psi-2\lambda[\phi,\psi]\\
&\qquad\qquad+
N\big(\epsilon_{ijk}(A^i\partial^jA^k+\tfrac13A^iA^jA^k)
+i\chi\chi\big)\Big)\ .
\label{av}
}
The theory has $\N=1$ supersymmetry (2 supercharges) and the field are
organized in to two multiplets, $(A_i,\chi_\alpha)$ and
$(\phi,\psi_\alpha)$. As we have already said, 
this theory will be subject to all kinds of
corrections coming from stringy effects and it is far form clear
whether it has any range of validity at all. In particular, in
\cite{Armoni:2005sp} we
argued that the topological mass scale $m=g^2N$ set by the level of the
Chern-Simons term is of order the string scale, {\it i.e.\/}~the
renormalization group scale $\Lambda$ of the
bulk $SU(N)$ theory, and so there is no actually no mass
hierarchy between the fields in \eqref{av} and other modes of the open
string or corrections coming from the truncation of the 
Born-Infeld action to the Yang-Mills action. However, it seems
that the Acharya-Vafa theory by itself is useful at least for
calculating the multiplicities of domain walls. But this is an index
theory calculation and hence is probably immune to the stringy
corrections.\footnote{Although, even here there is an interesting
  puzzle for the theory with gauge group $SO(N)$; see \cite{NEW}.} 
Our attitude is that, baring some unexpected miracle, 
the Acharya-Vafa theory will not, by itself,
describe the interactions between domain walls exactly, however, it is
worth investigating to see whether it has the right qualitative features.

\begin{figure}[ht]
\centerline{\includegraphics[width=2.5in]{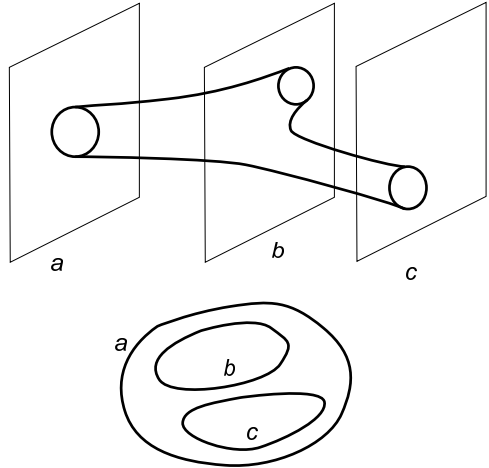}}
\caption{\footnotesize A thickened two-loop graph 
and the associated open string ``pants
  diagram'' involving three walls. } \label{triples}
\end{figure}

As usual, string perturbation theory corresponds to the large $N$
expansion in the $SU(N)$ gauge theory since $g_s\sim1/N$. The $1/N$
expansion corresponds to perturbation theory on the walls since the
wall coupling constant $g^2\sim\Lambda/N$ \cite{Armoni:2005sp}. String perturbation theory
therefore corresponds to perturbation theory of the domain wall theory. 
At leading
order, $g^0\sim 1/N^0$ we have the annulus diagram which vanishes because of
supersymmetry. This means that the leading order interaction
(in $1/N$) interaction term between $n$
domain walls is $\sim n^3$. This is counter-intuitive since, naively, one
would expect a binary 
interaction giving a dependence $\sim n^2$. 
The resolution of the puzzle is that the first
non-vanishing diagram is the ``pants diagram''
in Fig.~\eqref{triples}. 
In particular, it is clear that at this order, in the stringy picture, 
interactions involve either pairs, $a=c\neq b$, or triples, $a$, $b$
and $c$ all distinct, of walls. More
generally, at order $1/N^p$ interactions can involve at most $p+2$
walls. It augurs well for an underlying string theory that this 
dovetails with the binding energy per unit area of 
$n$ walls that follows from the exact BPS tension formula \eqref{tk}
\EQ{
\Delta T_n=nT_1-T_n=\frac{N^2\Lambda^3}{4\pi^2}\Big(n\sin\frac\pi N-
\sin\frac{\pi n}{N}\Big)
=\frac{\Lambda^3}{4}\sum_{p=1}^\infty (-1)^{p+1}
\frac{n(n^{2p}-1)}{(2p+1)!}\left(\frac\pi N\right)^{2p-1}\ .
\label{delta}
} 
The $n$ dependence of the term at order $1/N^{2p-1}$ is
\EQ{
\frac{n(n^{2p}-1)}{(2p+1)!}=\sum_{j=2}^{2p+1}c_j\MAT{n\\ j}\ ,
\label{ndep}
}
which can be interpreted as a sum over contributions from groups of
$j$ walls where $2\leq j\leq 2p+1$. 
Let us focus on the case leading
$1/N$ term ($p=1$). In this case we have ${n^3-n \over 6} =
{n(n-1)(n-2)\over 6} + {n(n-1) \over 2}$. The interpretation is that
there are two classes of contributions from the ``pants diagram''
Fig.~\eqref{triples}: one where the three strings 
end on different walls and one where two strings end on
one all and the remaining one ends on a different wall. The exact
tension formula implies that the relative weighting of these two
contributions is one-to-one. It is remarkable that the 
calculation presented in the next section, actually
confirms the above expectation: the various field theory contributions
sum to a potential whose
asymptotic behaviour predicts an $n^3-n$ dependence at the order $1/N$.

Another puzzle that follows from \eqref{delta} is why 
the expansion is in odd powers of $1/N$. In string theory D-branes
interactions involve all powers of $g_s$ and not only odd powers of
$g_s$. Due to the similarity between domain walls and D-branes we
expect that the interaction potential between domain walls
would include odd {\it and even\/} powers of $1/N$. However, the exact
tension \eqref{delta} exhibits only odd powers of $1/N$. Logically, it
is possible that the interaction {\it does\/} include even powers of
$1/N$ and only the tension does not. However, this seems
somewhat Machiavellian; in addition, as was already mentioned, the explicit
computation shows that the zeroth order contribution to the potential
vanishes identically. This is a hint that all the even power
contributions vanish identically due to some symmetry. 
This puzzle deserves further investigation.   

In the next section, we calculate the potential between $n$ walls at
order $1/N$ via a two-loop calculation in the domain wall theory.
This is generalization of the calculation in \cite{Armoni:2005sp} 
for the case of two walls.

\section{The Multi-Wall Potential at 2-Loops}

In this section, we consider the interactions between a set of $n$
parallel domain walls with arbitrary separations $X_a$. From the
point-of-view of the theory on the domain wall, we are at an arbitrary
point on the Coulomb branch. In the quantum theory, we expect that
this branch is lifted and that there will be forces between the
walls. From the domain wall perspective these forces appear as a
non-trivial Coleman-Weinberg effective potential on the Coulomb
branch once the massive degrees-of-freedom have been integrated out. 

On the Coulomb branch, the $U(n)$ gauge symmetry is broken to
$U(1)^n$. The overall $U(1)$ is completely decoupled form the
remaining degrees-of-freedom and so we can ignore it and work with a
$SU(n)$ theory instead. After symmetry breaking, 
\EQ{
\phi=\MAT{\varphi_1 & &\\ & \ddots & \\ &&\varphi_n}\ ,\qquad
\varphi_a\sim \Lambda^2X_a\ ,
}
only the diagonal
components of $\phi$ and $\psi$ are massless, all the other fields
either gain a mass through the Higgs mechanisms, or have a topological
mass coming from the Chern-Simons term, 
or a combination of a Higgs and topological mass. We discuss
them in turn below:

{\bf (1) Gauge bosons.} The off-diagonal components $A^{ab}_i$, $a\neq b$, 
charged under the a pair of unbroken $U(1)$'s,
have a complicated propagator which reflects a mixture between the Higgs
effect and the topological mass  
arising from the Chern-Simons
term.\footnote{A good reference for Yang-Mills-Chern-Simons theories
  is the review \cite{Dunne:1998qy}.} 
In Euclidean space, which we now use throughout, and Landau
gauge, the propagator is
\EQ{
\Delta_{ij}^{ab}(p)=\frac{(\delta_{ij}-p_ip_j/p^2)(p^2+\varphi_{ab}^2)-
m\epsilon_{ijk}p_k}{(p^2+m^{(+)2}_{ab})
(p^2+m_{ab}^{(-)2})}\ ,
\label{propa}
}
We have introduced the notation
$\varphi_{ab}\equiv\varphi_a-\varphi_b$ and defined the masses
\EQ{
m^{(\pm)}_{ab}=\sqrt{\varphi_{ab}^2+m^2/4}\pm m/2\ .
}
The diagonal components, neutral under the unbroken gauge group, 
$A^{aa}_i$ only have a topological mass. The propagator is still 
given by \eqref{propa} since $\varphi_{aa}=0$ and $m^{(\pm)}_{aa}=m$.

{\bf (2) Scalars.} The neutral components $\phi^{aa}$ are the massless Higgs
 fields while $\phi^{ab}$ are the would-be Goldstone Bosons and so 
are massless in Landau gauge.

{\bf (3) Fermions.} It is convenient to amalgamate the 
two 2-component fermion fields which are off-diagonal in colour indices 
into a 4-component field:
\EQ{
\Psi^{ab}=\MAT{\chi^{ab}_\alpha\\ \psi^{ab}_\alpha}\ .
}
In Euclidean space, the inverse propagator can 
then be written in $2\times 2$ block
form\footnote{Here, $\sigma_i=(\tau^1,\tau^2,\tau^3)$.}
\EQ{
\big(\Delta^{ab}_F(p)\big)^{-1}=\MAT{m-ip\cdot\sigma &
  i\varphi_{ab} \\ -i\varphi_{ab} & -ip\cdot\sigma}\ .
}
The remaining massive fields $\chi^{aa}_\alpha$ have mass $m$.

In addition to these fields and their interactions, we have to add the
usual gauge fixing terms and associated ghosts. The
vertices are those of a conventional spontaneously broken gauge theory
except that the
Chern-Simons term \eqref{av} modifies the momentum dependence of the 
three gauge vertex to
\EQ{
(p_1-p_2)_k\delta_{ij}+(p_2-p_3)_i\delta_{jk}+(p_3-p_1)_j\delta_{ik}
-m\epsilon_{ijk}
}
in Euclidean space. Note that in Euclidean space the
Chern-Simons term is pure imaginary.

The effective potential as a function of the VEVs $\varphi_a$ 
(which become the
field of the low-energy effective action) is obtained by integrating 
out all the massive modes: that is every field except $\phi^{aa}$ and
$\psi^{aa}$. In perturbation theory, the contribution is given by summing
all the vacuum graphs with massive fields propagating in the loops. It is
straightforward to verify that the one-loop contribution vanishes
identically due to the mass degeneracies entailed by
supersymmetry. At the two loop level, there are two kinds of vacuum
graph; namely, the sunset
and the figure-of-eight. Each sunset graph involves three
particles with charges $(ab)$, $(bc)$ and $(ac)$, while a
figure-of-eight involves two particles with charges $(ab)$ and
$(ac)$. Once the diagrams are thickened out to become open string
diagrams they both have the same topology and $a$, $b$ and $c$ become
Chan-Paton factors associated to three domain walls as illustrated in 
Figure \eqref{triples}. Although the theory
is finite, each separate graph is divergent and must be
regularized. Since we wish to preserve supersymmetry we use the dimensional
reduction regularization scheme where loop momenta
  propagate in $d$ dimensions, while the tensor and spinor structure
  is appropriate to 3 dimensions. If the
diagrams are calculated correctly, the poles in $d-3$ cancel due to
supersymmetry to leave a finite result. It should also be possible to do
the calculation in a real superspace formalism.

The loop integrals can all be calculated using the algorithm explained
in the Appendix of \cite{Armoni:2005sp}.
The contributions from individual diagrams are in general very
lengthy and since they have no real intrinsic meaning 
on their own we do not write down their contributions explicitly. 
However, once added together, enormous
cancellations and simplifications occur and for this reason will only
quote the end result for the effective potential:
\EQ{
V_\text{2-loop}(\varphi_a)=\frac{g^2m^2}{16\pi^2}
\sum_{a=1}^n\sum_{b=1}^n\sum_{c=1}^n
\frac{\varphi_{ab}\varphi_{cb}}{\sqrt{(4\varphi_{ab}^2
+m^2)(4\varphi_{bc}^2+m^2)}}\ .
\label{res}
}
It is worth emphasizing that the result depends on a sum over triplets of
$U(n)$ indices since, as we have already pointed out, 
individual diagrams involve at most three different Chan-Paton factors.

\section{Discussion}

Our result for the potential can be written in terms of the positions
of the walls in the transverse space by using 
$X_a\sim\varphi_a/\Lambda^2$. 
We can immediately cross check our result with the two wall
calculation in \cite{Armoni:2005sp}. The result 
\eqref{res} can naturally be written as a sum of a part 
which involves triples, $(a,b,c)$ all distinct, and
pairs, $a=c\neq b$. For a pair $a\neq b$, we have a contribution
\EQ{
\frac{g^2m^2}{8\pi^2}\frac{\varphi_{ab}^2}{m^2+4\varphi^2_{ab}}\ ,
}
which agrees with the two wall result of \cite{Armoni:2005sp}. 

The result \eqref{res} can also be written in terms of a (real)
superpotential $W$ as
\EQ{
V(\varphi_a)=
\sum_{a=1}^n\Big(\frac{\partial W}{\partial\varphi_a}\Big)^2\ ,
}
where
\EQ{
W_\text{2-loop}(\varphi_a)=\frac{2gm}\pi
\sum_{a=1}^n\sum_{b=1}^n\big(\sqrt{4\varphi_{ab}^2+m^2}-
m\big)=\frac{4gm}\pi\sum_{a=1}^n\sum_{b=1}^n m^{(-)}_{ab}\ .
\label{res2}
}
which makes it clear that there can be a supersymmetric
bound state when all the walls are coincident but no where else.
 
In the limit of small separation, $|\varphi_{ab}|\ll m$, we have
\EQ{
V_\text{2-loop}\longrightarrow\frac{g^2n}{32\pi^2}\sum_{a=1}^n\sum_{b=1}^n
\varphi_{ab}^2\ ,
}
which gives the mass scale $g^2\sim\Lambda/N$ for fluctuations around
the critical point. Notice that this is small, by a factor of $1/N$
than the string scale $\Lambda$, and so should be a robust prediction from the
truncated theory. 
In the other limit, of large separations, $|\varphi_{ab}|\gg m$, we have
\EQ{
V_\text{2-loop}\longrightarrow\frac{g^2m^2}{32\pi^2}\cdot\frac{n^3-n}6\ .
\label{asl}
}
There are two remarkable things about this result: that it is a
constant at all and that the result has the correct $n$
dependence to match \eqref{ndep}. This is very unexpected since it
means that the relative weighting between the contributions from triples of
walls and pairs of walls is exactly right. This is puzzling because it
is clear that the power law behaviour of the potential for large
separations cannot be physical since there is a mass gap in the 4d theory.
Hence, the large distance potential must be of a Yukawa type. 
 It seems
plausible that for large separations stringy effects could change 
the power law to exponential
behaviour; however, why should this not also effect the constant and
destroy the $n$ dependence?
Note that \eqref{asl} also has the correct $N$ and
$\Lambda$ dependence once one uses the substitutions $m=g^2N$ and
$g^2\sim\Lambda/N$.

In summary: we have calculated the domain wall potential at order
$1/N$ from thinking of them as D-branes for the confining string. 
We expect on general grounds that the result can only really be trusted 
at sub-stringy distances, $X \ll \Lambda ^{-1}$, and, in particular, we
found a result that was consistent with the existence of a
supersymmetric bound state of $n$ walls. What is surprising is that 
we found a result
that seems to be consistent also at very large distances: why?

\vspace{1cm}

{\bf Acknowledgments:}  A.A. is supported by a PPARC Advanced
Fellowship.

\end{document}